\documentclass[prb,amsmath,amssymb,superscriptaddress,twocolumn]{revtex4}
\usepackage{braket}
\usepackage{graphicx}
\usepackage{bbm,color,ulem}
\DeclareMathAlphabet{\mathpzc}{OT1}{pzc}{m}{it}
\begin{document}
\title{Conditions allowing error correction in driven qubits}
\author{Robert E.\ Throckmorton}
\author{S.\ Das Sarma}
\affiliation{Condensed Matter Theory Center and Joint Quantum Institute, Department of Physics, University of Maryland, College Park, Maryland 20742-4111 USA}
\date{\today}
\begin{abstract}
We consider a qubit that is driven along its logical $z$ axis, with noise along the $z$ axis
in the driving field $\Omega$ proportional to some function $f(\Omega)$, as well as noise along
the logical $x$ axis.  We establish that whether or not errors due to both types of noise can
be canceled out, even approximately, depends on the explicit functional form of $f(\Omega)$ by
considering a power-law form, $f(\Omega)\propto\Omega^k$.  In particular, we show that such
cancellation is impossible for $k=0$, $1$, or any even integer.  However, any other odd integer
value of $k$ besides $1$ does permit cancellation; in fact, we show that both types of errors
can be corrected with a sequence of four square pulses of equal duration.  We provide sets of
parameters that correct for errors for various rotations and evaluate the error, measured by
the infidelity, for the corrected rotations versus the na\"ive rotations, i.e., the operations
that, in the complete absence of noise, would produce the desired rotations (in this case a
single pulse of appropriate duration and magnitude).  We also consider a train of four
trapezoidal pulses, which take into account the fact that there will be, in real experimental
systems, a finite rise time, again providing parameters for error-corrected rotations that
employ such pulse sequences.  Our dynamical decoupling error correction scheme works for any
qubit platform as long as the errors are quasistatic. 
\end{abstract}
\maketitle

\section{Introduction}
Error correction is one of the most important topics in the field of quantum computation.
In fact, the whole subject of quantum computation as a practical field exists simply because
quantum error correction is theoretically possible provided the error is not too large.
Unlike the bits in a classical computer, which can only take one of two states, $0$ and $1$,
a qubit in a quantum computer can take on an uncountably infinite number of different states
of the form, $\alpha\ket{0}+\beta\ket{1}$, where $\alpha$ and $\beta$ are complex numbers
satisfying $|\alpha|^2+|\beta|^2=1$. As a result, even small errors in the state of a qubit
will adversely affect the results of a computation.  Since decoherence-related errors arising
from external noise in the environment (even in the absence of any explicit errors in gate
operations) are unavoidable in a quantum system, carrying out practical quantum error corrections
is the key roadblock in building effective quantum computing circuits.  The development of
techniques for correcting for errors is therefore of utmost importance.  In fact, much research has been devoted
to just this topic, taking various approaches.  One such approach is to build in error resistance
via engineering of the physical system and error correction using ancilla qubits.  For example,
in semiconductor-based electron spin qubits, magnetic noise due to the presence of
magnetic isotopes (i.e., non-zero nuclear spin in the environment) is a major source of error.  This may be
reduced by such techniques as isotopic purification (if possible) and polarization of the atomic
nuclei.  Another important error in semiconductor (and other solid state) qubits is the charge noise
in the environment.  One method that employs ancilla qubits to perform error correction is the surface code
technique\cite{FowlerPRA2012}.  Topological qubits, such as the Majorana qubit, are, to an extent,
immune to error\cite{DasSarmaNPJQI2015} (i.e., some, though not all, operations are protected).  Error cancellation methods
that use designed pulses include the NMR-inspired Hahn echo method and its generalization, the
Carr-Purcell-Meiboom-Gill (CPMG) technique\cite{WitzelPRL2007,WitzelPRB2007,LeePRL2008},
Bayesian estimation of parameters\cite{ShulmanNatCommun2014,SergeevichPRA2011}, and dynamical
decoupling through specially-designed pulse sequences\cite{ShulmanScience2012,BluhmPRL2010,BluhmNatPhys2011,SergeevichPRA2011,MuhonenNatNanotechnol2014,MalinowskiNatNanotechnol2017}.
In fact, carrying out error correction protocols through various alternative techniques, employing
both software and hardware methods, is the single most active area in the subject of quantum computation,
and progress in general has been slow in the sense that no qubit platform has yet experimentally
achieved one single error-free logical qubit.

Our current work will focus on error correction via pulse engineering through the dynamical decoupling
technique.  We generalize the recent work of Ref.\ \onlinecite{ZengNJP2018}, which considered a qubit
driven only along the logical $z$ axis subject to a noise term along its logical $x$ axis (orthogonal-to-driving-field
noise), with a Hamiltonian of the form,
\begin{equation}
H=\tfrac{1}{2}\Omega(t)\,\sigma_z+\delta\beta\,\sigma_x,
\end{equation}
where $\Omega(t)$ is the driving field and the $\delta\beta$ is a noise-induced error
term.  We summarize below the main finding of Ref.\ \onlinecite{ZengNJP2018} relevant
for our work.  In Ref.\ \onlinecite{ZengNJP2018}, which itself is a special case of
the general geometric method for dynamical decoupling developed in Ref.\ \onlinecite{BarnesSciRep2015},
it is shown that the conditions for correcting for this error term along the $x$ axis
(with the control driving field along the $z$ axis) without any error in the driving
field may be cast into a geometric picture; specifically, it is shown that the condition
that results in cancellation of errors due to $\delta\beta$ to first order is equivalent
to the condition that the curve traced out in the complex plane by
\begin{equation}
x(t)+iy(t)=\int_{0}^{t}e^{i\phi(t')}\,dt',
\end{equation}
where $\phi(t)$, the rotation angle at time $t$, given by
\begin{equation}
\phi(t)=\frac{1}{\hbar}\int_{0}^{t}\Omega(t')\,dt',
\end{equation}
must be closed, i.e.,
\begin{equation}
\int_{0}^{T}e^{i\phi(t)}\,dt=0, \label{Eq:DBCondition}
\end{equation}
where $T$ is the duration of the gate.  The driving field is simply proportional to the
curvature of this curve,
\begin{equation}
\frac{\Omega(t)}{\hbar}=\frac{\dot{x}\ddot{y}-\ddot{x}\dot{y}}{(\dot{x}^2+\dot{y}^2)^{3/2}}.
\end{equation}
Conditions for cancellation of higher-order terms are also found, imposing further
restrictions on this curve.  We will generalize these results to the case in which
noise in the driving field is also present, i.e., we consider the Hamiltonian,
\begin{equation}
H=\tfrac{1}{2}[\Omega(t)+f(\Omega)\,\delta\epsilon]\sigma_z+\delta\beta\,\sigma_x, \label{Eq:Hamiltonian}
\end{equation}
where $f(\Omega)$ is some function of the driving field.  In this case, $\delta\epsilon$
represents fluctuations in some parameter that controls the driving field.  We refer the
reader to Ref.\ \onlinecite{ZengNJP2018} for the details and the background on the geometric
methods for dynamical decoupling and for a general review of the literature in this context,
focusing entirely in our work on how to correct for errors existing along both the $x$ and $z$ axes
as shown explicitly in Eq.\ \eqref{Eq:Hamiltonian} above.  Throughout this work, we will
make the quasistatic approximation, i.e., we will assume that $\delta\epsilon$ and $\delta\beta$
remain constant for the entirety of a given gate.  Thus, the applied pulses are thought to
be fast compared with the dynamical time scale of the error terms in Eq.\ \eqref{Eq:Hamiltonian},
which is a reasonable approximation for many types of ``slow noise'' arising in practical qubits.
Also, this ``slow-noise'' approximation could always be satisfied, in principle, by making
the external pulses and drives faster.  We will also be considering power-law forms for $f(\Omega)$, i.e., $f(\Omega)\propto\Omega^k$.
Our main goal is to determine whether or not correction of both types of errors (i.e., $\delta\beta$
and $\delta\epsilon$) is possible for a given value of $k$.  Note that the errors themselves
are considered to be completely unknown and arbitrary except for their quasistatic nature,
i.e., the ``slow-noise'' assumption.

While our results are completely general and make no reference to a specific qubit platform
or architecture, as long as it can be represented by a Hamiltonian of this form, Eq.\ \eqref{Eq:Hamiltonian},
we will give a few examples of physical systems that it could represent.  The simplest example
of such a system would be a single electron spin in a tunable magnetic field along the $z$ axis;
here, $\Omega(t)$ would be the Zeeman energy due to this field\cite{KoppensNature2006}.  The
error terms would then represent fluctuations in this field.  Note that the two errors, $\delta\beta$
and $\delta\epsilon$, represent respectively the fluctuations in the transverse and longitudinal
fields, both being of crucial importance in quantum computing.  Another such example is a singlet-triplet
qubit\cite{PettaScience2005} with no intentionally-applied magnetic field gradient.  In this
case, $\Omega(t)$ is the (intended) exchange coupling between the two spins, which is controlled
by changing the detuning $\epsilon$ between the two quantum dots; $\delta\epsilon$ then represents
the noise in the detuning.  One may also encounter magnetic noise that induces a field gradient,
given by the $\delta\beta$ term.  We emphasize, however, that our theory is completely general and
applies to all driven qubits obeying Eq.\ \eqref{Eq:Hamiltonian} under the slow noise approximation
(i.e., time-independent $\delta\beta$ and $\delta\epsilon$).

To give a concrete example of the energy and time scales of error-inducing terms in this particular
platform, we consider the experimental data and analysis thereof of Ref.\ \onlinecite{BarnesPRB2016}.
The standard deviation of the fluctuations in the magnetic field gradient is found to be about $23\text{ MHz}$,
or $95.1\text{ neV}$.  The corresponding standard deviation in the exchange coupling is found to be
about $4.26\times 10^{-3}\Omega$, implying an approximately linear relation for $f(\Omega)$ in this
case.  However, this linear relation is not expected to hold exactly, and in fact breaks down precisely
in the regime within which these qubits are typically operated\cite{ButerakosArXiv}.  As for time
scales, the noise fluctuations occur over microsecond time scales, while gates are performed over
nanosecond time scales.  As stated earlier, the faster we make our gates, the larger the absolute
magnitude of the pulses need to be, so that the assumption that the errors are small and quasistatic
will be justified.  We should emphasize that these numbers are really only relevant to the specific
device considered; the numbers may differ for other devices and physical realizations of the qubit.

We start by solving for the time evolution operator for the above Hamiltonian, Eq.\ \eqref{Eq:Hamiltonian},
treating the error terms perturbatively.  We then simply require that the first-order terms in
$\delta\beta$ and $\delta\epsilon$ be zero.  The condition that we obtain for canceling
errors due to the $\delta\beta$ term is identical to that found in Ref.\ \onlinecite{ZengNJP2018},
and thus the geometric picture described therein and summarized above applies here as well.
The condition that results in the cancellation of the $\delta\epsilon$ term,
\begin{equation}
\int_{0}^{T}f[\Omega(t)]\,dt=0, \label{Eq:DECondition}
\end{equation}
is just a restriction on the curvature of the curve.  If the driving field noise is the {\it only}
type of noise present in the system, it may easily be shown that this condition will in fact {\it exactly}
cancel this noise.  Applied to the power-law form of $f(\Omega)$, we will see that we can cast
all of the conditions into forms that are independent of the proportionality constant in $f(\Omega)$.
Applying this condition, we will find that, if $k=0$ or 1, then full error correction for arbitrary
rotations is impossible---in the former case, it is not possible to correct {\it any} rotation, and
in the latter, only the identity operation, for which $\phi=0$, can be corrected.  This comes from
the fact that the integral is just proportional to $\phi$ when $k=1$.  Furthermore, even values of
$k$ do not allow for this condition to be satisfied; the integrand on the left-hand side will always
be non-negative, and thus the integral can only be zero if $\Omega(t)=0$ for all times $t$.  The only
values of $k$ for which error correction may be possible are therefore odd values other than $1$.  We
show, in fact, that error correction {\it is} possible for these cases.  This is simply due to the fact
that Eq.\ \eqref{Eq:DECondition} yields a distinct condition on the driving field from the condition
that the operation result in a rotation by a specific angle $\phi$ about the $z$ axis,
\begin{equation}
\frac{1}{\hbar}\int_{0}^{T}\Omega(t)\,dt=\phi,
\end{equation}
rather than a redundant ($k=1$ and $\phi=0$) or contradictory ($k=0$ or $1$ and $\phi\neq 0$)
condition.  We will consider the case of four square pulses of equal duration.  Our choice of
this sequence comes from the fact that there are four real-valued conditions, along with recent
results indicating that the fastest pulses that cancel errors due to the $\delta\beta$ term are
square pulses\cite{ZengArXiv}.  The four real-valued conditions are just the two error-cancellation
conditions, Eqs.\ \eqref{Eq:DBCondition} and \eqref{Eq:DECondition}, along with the simple
requirement that the gate perform a rotation about the $z$ axis by a specific angle.  We
provide parameters for this four-pulse sequence that cancel both types of error for $k=-3$,
$-1$, $3$, and $5$, though similar parameters can be found for other suitable values of $k$
as well.

While these sequences made up of square pulses provide a proof of principle demonstrating that
error correction is possible for the forms of $f(\Omega)$ considered, such perfect square pulses
are not realizable in actual experiments.  They assume that the driving field can be increased
arbitrarily rapidly, which is not the case---there is always a finite rise time.  As a result,
we are interested in considering pulses that include such finite rise times for a practical
implementation of our proposal.  In particular, we consider a train of trapezoidal pulses, which
consist of a linear rise or fall, followed by a constant segment.  This choice is motivated by
the fact that, for such a pulse sequence, the equations giving the constraints, while complicated,
can be obtained in analytic form, albeit in terms of non-elementary functions (Fresnel integrals).
We consider a train of four such pulses, plus a final ramp down to $\Omega=0$, and repeat the above
analysis, this time only for $f(\Omega)\propto\Omega^3$.  We once again show that error correction
can be done, this time for a more realistic pulse sequence.

The rest of this article is organized as follows.  We derive the general conditions for
cancellation of errors to first order in Sec.\ II.  We then specialize to the case of
four square pulses in Sec.\ III and that of four trapezoidal pulses in Sec.\ IV, providing
pulse parameters for each case and evaluating the performance of these pulses in terms of
the fidelity of the operations.  Finally, we give our conclusions in Sec.\ V.

\section{Error cancellation conditions}
Let us begin by deriving the error cancellation conditions, Eqs.\ \eqref{Eq:DBCondition} and
\eqref{Eq:DECondition}.  We will consider here a qubit with the Hamiltonian given in Eq.\ \eqref{Eq:Hamiltonian}.
We make the quasistatic approximation, i.e., we assume that $\delta\beta$ and $\delta\epsilon$
are independent of time.  Furthermore, we assume that the error in the driving field is
proportional to some function $f(\Omega)$ of said driving field.

Without the error terms, we can solve for the evolution operator exactly; it is simply
\begin{equation}
U_0(t)=\left [\begin{matrix}
e^{-i\phi(t)/2} & 0 \\
0 & e^{i\phi(t)/2}
\end{matrix}\right ],
\end{equation}
where
\begin{equation}
\phi(t)=\frac{1}{\hbar}\int_{0}^{t}\Omega(t')\,dt'.
\end{equation}

We now add back in the error terms and attempt a perturbative solution in powers of $\delta\beta$
and $\delta\epsilon$.  We decompose the full time evolution operator into two parts, $U(t)=U_0(t)U_p(t)$.
If we substitute this into the time-dependent Schr\"odinger equation for the system, we find that
\begin{equation}
i\hbar\frac{dU_p}{dt}=H_{\text{eff}}U_p,
\end{equation}
where
\begin{eqnarray}
H_{\text{eff}}&=&U_0^{\dag}[H-\tfrac{1}{2}\Omega(t)\sigma_z]U_0 \cr
&=&\tfrac{1}{2}f(\Omega)\sigma_z\,\delta\epsilon+[\cos{\phi(t)}\sigma_x-\sin{\phi(t)}\sigma_y]\,\delta\beta.
\end{eqnarray}
Because $U_p(t)$ is a $2\times 2$ unitary operator, it may be written in the general form,
\begin{equation}
U_p(t)=\left [\begin{matrix}
u & -v^{\ast} \\
v & u^{\ast}
\end{matrix}\right ].
\end{equation}
If we now substitute this into the effective Schr\"odinger equation, we obtain the following equations:
\begin{eqnarray}
i\hbar\frac{du}{dt}&=&\tfrac{1}{2}f(\Omega)u\,\delta\epsilon+e^{i\phi}v\,\delta\beta, \\
i\hbar\frac{dv}{dt}&=&-\tfrac{1}{2}f(\Omega)v\,\delta\epsilon+e^{-i\phi}u\,\delta\beta.
\end{eqnarray}
We now expand $u$ and $v$ in power series in $\delta\beta$ and $\delta\epsilon$:
\begin{eqnarray}
u(t)&=&\sum_{k,l=0}^{\infty}g_{kl}(t)(\delta\epsilon)^k(\delta\beta)^l, \\
v(t)&=&\sum_{k,l=0}^{\infty}h_{kl}(t)(\delta\epsilon)^k(\delta\beta)^l.
\end{eqnarray}
If we substitute these expansions into the above equations and equate like powers, we obtain the recursion
relations,
\begin{eqnarray}
i\hbar\frac{dg_{kl}}{dt}&=&\tfrac{1}{2}f(\Omega)g_{k-1,l}+e^{i\phi}h_{k,l-1}, \\
i\hbar\frac{dh_{kl}}{dt}&=&e^{-i\phi}g_{k,l-1}-\tfrac{1}{2}f(\Omega)h_{k-1,l}.
\end{eqnarray}
The initial conditions are $g_{00}(t)=1$ and $h_{00}(t)=0$.  Furthermore, $g_{kl}(t=0)=h_{kl}(t=0)=0$ if at
least one of $k$ and $l>0$.

We now apply a pulse of duration $T$ to the system.  We want to find the conditions under which this
pulse will have zero error to first order in $\delta\epsilon$ and $\delta\beta$.  If we now write down the
equations giving these first-order corrections, we get
\begin{eqnarray}
i\hbar\frac{dg_{10}}{dt}&=&\tfrac{1}{2}f(\Omega), \\
i\hbar\frac{dh_{10}}{dt}&=&0, \\
i\hbar\frac{dg_{01}}{dt}&=&0, \\
i\hbar\frac{dh_{01}}{dt}&=&e^{-i\phi}.
\end{eqnarray}
If we now solve these equations, we find that $g_{01}(t)=h_{10}(t)=0$, while
\begin{eqnarray}
g_{10}(t)&=&-\frac{i}{2\hbar}\int_{0}^{t}f[\Omega(t')]\,dt', \\
h_{01}(t)&=&-\frac{i}{\hbar}\int_{0}^{t}e^{-i\phi(t')}\,dt'.
\end{eqnarray}
Since $g_{01}(t)$ and $h_{10}(t)$ are already identical to zero for all times $t$, the condition for the
cancellation of error due to orthogonal-to-driving-field noise to first order, represented by the $\delta\beta$
term, is
\begin{equation}
\int_{0}^{T}e^{-i\phi(t)}\,dt=0,
\end{equation}
while that for driving field noise, represented by the $\delta\epsilon$ term, is
\begin{equation}
\int_{0}^{T}f[\Omega(t)]\,dt=0.
\end{equation}
These are just Eqs.\ \eqref{Eq:DBCondition} and \eqref{Eq:DECondition}, respectively.  The first of these
conditions is the same as that which was previously obtained in Ref.\ \onlinecite{ZengNJP2018} in the case in which only
$\delta\beta$ was present, i.e., assuming no noise in the driving field.  The second simply imposes a
further condition on the driving field.  Note that, not unexpectedly, the existence of the longitudinal
noise $\delta\epsilon$ along the drive direction does not affect the condition for correcting the transverse
error $\delta\beta$.

We may show that, in fact, if $\delta\beta=0$, then Eq.\ \eqref{Eq:DECondition} will result in {\it exact}
(i.e., to all orders) cancellation of errors due to driving field noise.  To see this, we simply note that,
if $\delta\beta=0$, then we can solve for the evolution operator exactly, obtaining
\begin{equation}
U(t)=\left [\begin{matrix}
e^{-i\Phi(t)/2} & 0 \\
0 & e^{i\Phi(t)/2}
\end{matrix}\right ],
\end{equation}
where $\Phi(t)=\phi(t)+\delta\phi(t)$ and
\begin{equation}
\delta\phi(t)=\frac{1}{\hbar}\int_{0}^{t}f[\Omega(t')]\,dt'\,\delta\epsilon.
\end{equation}
We can thus cancel the error exactly if we require that $\delta\phi(T)=0$, which gives us Eq.\ \eqref{Eq:DECondition}.
Of course, we could also cancel the error by letting $\delta\phi(T)=2\pi n$, with $n$ any integer, but a
nonzero $n$ would require a precise knowledge of $\delta\epsilon$, thus making such error cancellation very
impractical.

We now consider two simple cases---$f(\Omega)=C$, where $C$ is an arbitrary dimensionless constant, and
$f(\Omega)=\frac{\Omega}{\Omega_0}$, where $\Omega_0$ is an arbitrary constant with units of energy.  In
the first case, if we substitute the form of $f(\Omega)$ into the left-hand side of Eq.\ \eqref{Eq:DECondition},
we get $CT$.  This condition can only be satisfied if $C=0$ or $T=0$, i.e., we either have no driving field
noise at all or we simply do not perform an operation.  If $C\neq 0$, we see that it is impossible to cancel the
error for {\it any} operation.  For the second case, in which the noise depends linearly on the driving
field, we find that the left-hand side of Eq.\ \eqref{Eq:DECondition} is just
\begin{equation}
\frac{1}{\Omega_0}\int_{0}^{T}\Omega(t)\,dt.
\end{equation}
This is just proportional to $\phi(T)$:
\begin{equation}
\frac{1}{\Omega_0}\int_{0}^{T}\Omega(t)\,dt=\frac{\hbar}{\Omega_0}\phi(T).
\end{equation}
This tells us that, to cancel the driving field error, we would need to be simply performing an identity
operation, i.e., $\phi(T)=0$.  Therefore, if the noise in the driving field is proportional to the driving
field, then it is not possible to devise a means to cancel the driving field noise, even just to first order,
for any non-trivial (i.e., other than the identity) operation.

We will show, however, that other forms of $f(\Omega)$ {\it do} allow for cancellation of driving field
noise for arbitrary operations.  One such form is a power-law dependence, $f(\Omega)=\left (\frac{\Omega}{\Omega_0}\right )^{2n+1}$,
with an odd power, i.e., for integer $n$.  We note that the power {\it must} be odd (and, as just shown,
not equal to $1$), since, if the power were even, then $f(\Omega)$ will always be non-negative, and thus
we cannot satisfy the driving field error cancellation condition, Eq.\ \eqref{Eq:DECondition}, unless
$\Omega(t)=0$ for all times $t$.

\section{Train of square pulses}
\begin{figure}[htb]
	\centering
		\includegraphics[width=\columnwidth]{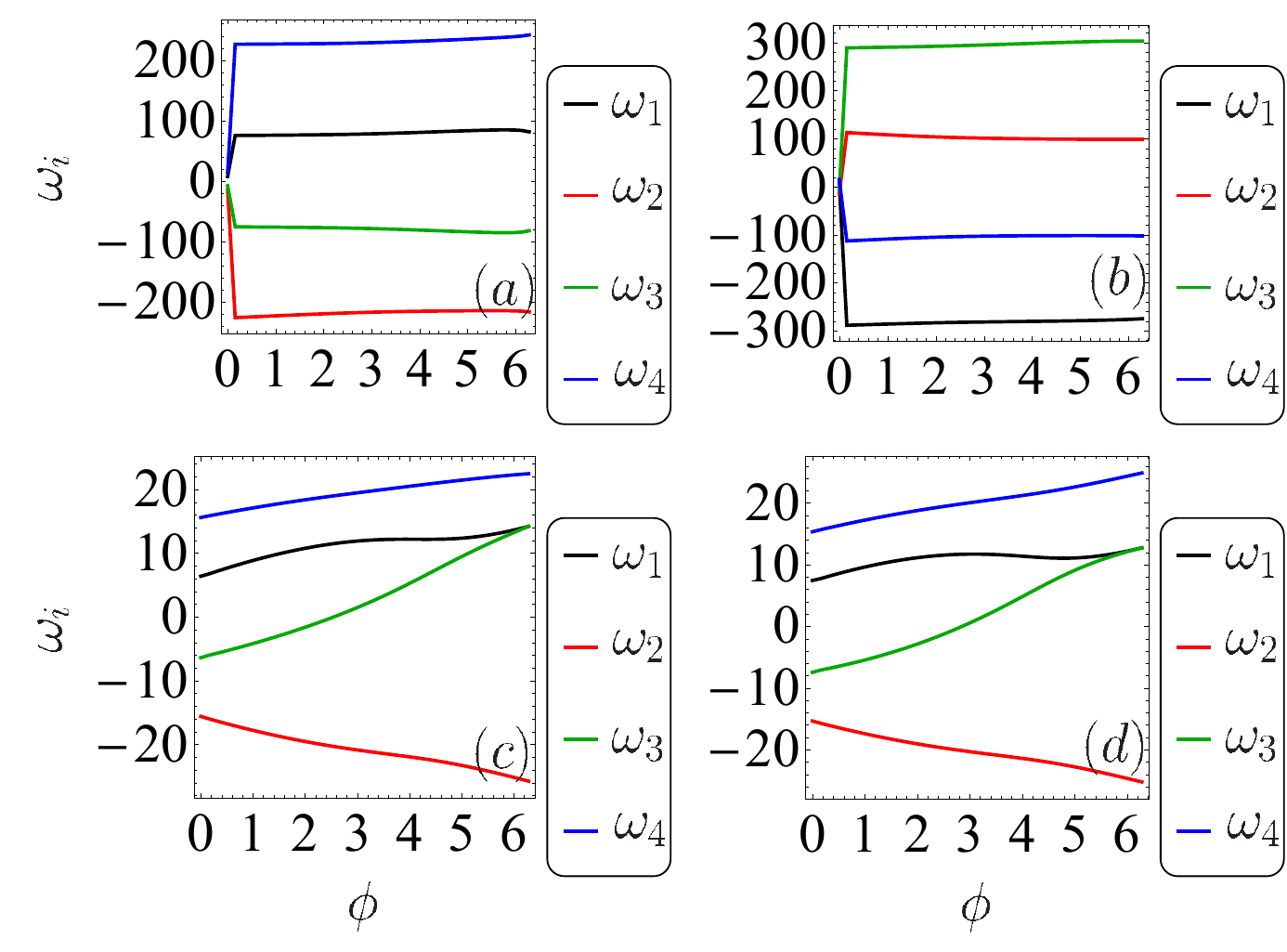}
		\caption{Plot of the dimensionless pulse paramaters $\omega_k$ as a function of the rotation angle $\phi$
		for $f(\Omega)\propto\Omega^{2n+1}$ and for $n=-2$ ($a$), $-1$ ($b$), $1$ ($c$), and $2$ ($d$).  These
		parameters are obtained by solving Eqs.\ \eqref{Eq:RotCondition}--\eqref{Eq:DBCancellationCond} numerically.}
	\label{Fig:Params_PowerLaw}
\end{figure}

We will demonstrate this by showing that a sequence of four square pulses can cancel both types of error
if $f(\Omega)$ is of such an odd (including negative integers) power-law form.  In fact, for such forms,
we will find that the parameters do not depend on the constant $\Omega_0$, but only on the power $2n+1$.
We require four pulses because we have four (real) conditions to satisfy in order to perform a rotation
by an angle $\phi$ and to cancel out error due to driving field and orthogonal-to-driving-field noise.
We will assume throughout, without any loss of generality, that the four pulses are of equal duration,
i.e., they each last for a time $\tfrac{1}{4}T$, and will label the value of the driving field for each
segment $\Omega_k$, with $k$ running from $1$ to $4$.  Doing this, the condition that we perform a rotation
by $\phi$ becomes
\begin{equation}
\frac{T}{4\hbar}(\Omega_1+\Omega_2+\Omega_3+\Omega_4)=\phi.
\end{equation}
Let us now introduce the dimensionless quantities, $\omega_k=\Omega_kT/\hbar$.  We may then write
\begin{equation}
\tfrac{1}{4}(\omega_1+\omega_2+\omega_3+\omega_4)=\phi. \label{Eq:RotCondition}
\end{equation}
Similarly, we may write the condition for canceling driving field noise as
\begin{equation}
\omega_1^{2n+1}+\omega_2^{2n+1}+\omega_3^{2n+1}+\omega_4^{2n+1}=0.
\end{equation}
Finally, the condition for canceling error from the $\delta\beta$ term is
\begin{eqnarray}
\frac{e^{-i\omega_1/4}-1}{\omega_1}&+&\frac{e^{-i\omega_1/4}(e^{-i\omega_2/4}-1)}{\omega_2} \cr
&+&\frac{e^{-i(\omega_1+\omega_2)/4}(e^{-i\omega_3/4}-1)}{\omega_3} \cr
&+&\frac{e^{-i(\omega_1+\omega_2+\omega_3)/4}(e^{-i\omega_4/4}-1)}{\omega_4}=0. \label{Eq:DBCancellationCond}
\end{eqnarray}
We will now need to solve for the values of the $\omega_k$ numerically.  We have obtained values for the
cases, $n=-2$, $-1$, $1$, and $2$, corresponding to $f(\Omega)\propto\Omega^{-3}$, $\Omega^{-1}$, $\Omega^3$,
and $\Omega^5$, respectively, and for arbitrary rotations, thus showing that it is in fact possible to
correct for both driving field noise and orthogonal-to-driving-field noise.  We present plots of the parameters
that we obtain below as a function of the rotation angle $\phi$ for $0\leq\phi\leq 2\pi$ in Fig.\ \ref{Fig:Params_PowerLaw},
as well as tables of values for some of these rotations in Tables \ref{Tab:PulseParameters_m2}--\ref{Tab:PulseParameters_2}.
These are by no means the only solutions to the equations given above; they are provided as examples to
show that it is in fact possible to satisfy all of the required constraints for the values of $n$ considered.
Also, we note that these parameters are independent of the proportionality constant, as it cancels out of
the equations.

\begin{table}[htb]
	\centering
		\begin{tabular}{c|c|c|c|c}
			$\phi$ & $\omega_1$ & $\omega_2$ & $\omega_3$ & $\omega_4$ \\
			\hline
			\hline
			$\pi/4$ & 75.8215 & -223.357 & -75.8627 & 226.54 \\
			\hline
			$\pi/2$ & 76.3572 & -220.639 & -76.4437 & 227.008 \\
			\hline
			$3\pi/4$ & 77.1698 & -218.215 & -77.3077 & 227.777 \\
			\hline
			$\pi$ & 78.4934 & -216.345 & -78.6919 & 229.11 \\
			\hline
			$5\pi/4$ & 80.3912 & -215.143 & -80.6635 & 231.124 \\
			\hline
			$3\pi/2$ & 82.5698 & -214.433 & -82.9295 & 233.642 \\
			\hline
			$7\pi/4$ & 84.4567 & -214.117 & -84.908 & 236.56 \\
			\hline
			$2\pi$ & 81.5261 & -216.137 & -81.9458 & 241.689
		\end{tabular}
		\caption{Pulse parameters for different rotation angles $\phi$ for $f(\Omega)\propto\Omega^{-3}$.  These
		parameters are obtained by solving Eqs.\ \eqref{Eq:RotCondition}--\eqref{Eq:DBCancellationCond} numerically.}
	\label{Tab:PulseParameters_m2}
\end{table}
\begin{table}[htb]
	\centering
		\begin{tabular}{c|c|c|c|c}
			$\phi$ & $\omega_1$ & $\omega_2$ & $\omega_3$ & $\omega_4$ \\
			\hline
			\hline
			$\pi/4$ & -286.42 & 109.271 & 290.092 & -109.801 \\
			\hline
			$\pi/2$ & -284.197 & 105.476 & 291.465 & -106.461 \\
			\hline
			$3\pi/4$ & -282.446 & 102.662 & 293.266 & -104.057 \\
			\hline
			$\pi$ & -281.248 & 100.883 & 295.603 & -102.672 \\
			\hline
			$5\pi/4$ & -280.291 & 99.7883 & 298.176 & -101.966 \\
			\hline
			$3\pi/2$ & -279.218 & 99.0761 & 300.637 & -101.646 \\
			\hline
			$7\pi/4$ & -277.66 & 98.6883 & 302.633 & -101.67 \\
			\hline
			$2\pi$ & -274.699 & 98.8635 & 303.306 & -102.337
		\end{tabular}
		\caption{Pulse parameters for different rotation angles $\phi$ for $f(\Omega)\propto\Omega^{-1}$.  These
		parameters are obtained by solving Eqs.\ \eqref{Eq:RotCondition}--\eqref{Eq:DBCancellationCond} numerically.}
	\label{Tab:PulseParameters_m1}
\end{table}
\begin{table}[htb]
	\centering
		\begin{tabular}{c|c|c|c|c}
			$\phi$ & $\omega_1$ & $\omega_2$ & $\omega_3$ & $\omega_4$ \\
			\hline
			\hline
			$\pi/4$ & 8.36239 & -17.3526 & -4.66872 & 16.8005 \\
			\hline
			$\pi/2$ & 10.0363 & -18.837 & -2.77338 & 17.8573 \\
			\hline
			$3\pi/4$ & 11.2749 & -20.0533 & -0.58396 & 18.7871 \\
			\hline
			$\pi$ & 11.9997 & -21.0243 & 1.96658 & 19.6244 \\
			\hline
			$5\pi/4$ & 12.1933 & -21.8642 & 4.95667 & 20.4222 \\
			\hline
			$3\pi/2$ & 12.226 & -22.8384 & 8.26825 & 21.1937 \\
			\hline
			$7\pi/4$ & 12.8317 & -24.1614 & 11.422 & 21.8989 \\
			\hline
			$2\pi$ & 14.2087 & -25.753 & 14.2087 & 22.4684
		\end{tabular}
		\caption{Pulse parameters for different rotation angles $\phi$ for $f(\Omega)\propto\Omega^3$.  These
		parameters are obtained by solving Eqs.\ \eqref{Eq:RotCondition}--\eqref{Eq:DBCancellationCond} numerically.}
	\label{Tab:PulseParameters_1}
\end{table}
\begin{table}[htb]
	\centering
		\begin{tabular}{c|c|c|c|c}
			$\phi$ & $\omega_1$ & $\omega_2$ & $\omega_3$ & $\omega_4$ \\
			\hline
			\hline
			$\pi/4$ & 9.1941 & -16.9967 & -5.90998 & 16.8542 \\
			\hline
			$\pi/2$ & 10.5926 & -18.3902 & -4.07217 & 18.153 \\
			\hline
			$3\pi/4$ & 11.4567 & -19.5346 & -1.75304 & 19.2557 \\
			\hline
			$\pi$ & 11.7197 & -20.4685 & 1.10502 & 20.2102 \\
			\hline
			$5\pi/4$ & 11.4261 & -21.3171 & 4.47574 & 21.1233 \\
			\hline
			$3\pi/2$ & 11.061 & -22.3193 & 7.94993 & 22.1579 \\
			\hline
			$7\pi/4$ & 11.4561 & -23.655 & 10.7573 & 23.4328 \\
			\hline
			$2\pi$ & 12.7367 & -25.2161 & 12.7367 & 24.8754
		\end{tabular}
		\caption{Pulse parameters for different rotation angles $\phi$ for $f(\Omega)\propto\Omega^5$.  These
		parameters are obtained by solving Eqs.\ \eqref{Eq:RotCondition}--\eqref{Eq:DBCancellationCond} numerically.}
	\label{Tab:PulseParameters_2}
\end{table}

\subsection{Evaluation of error}
\begin{figure}[htb]
	\centering
		\includegraphics[width=\columnwidth]{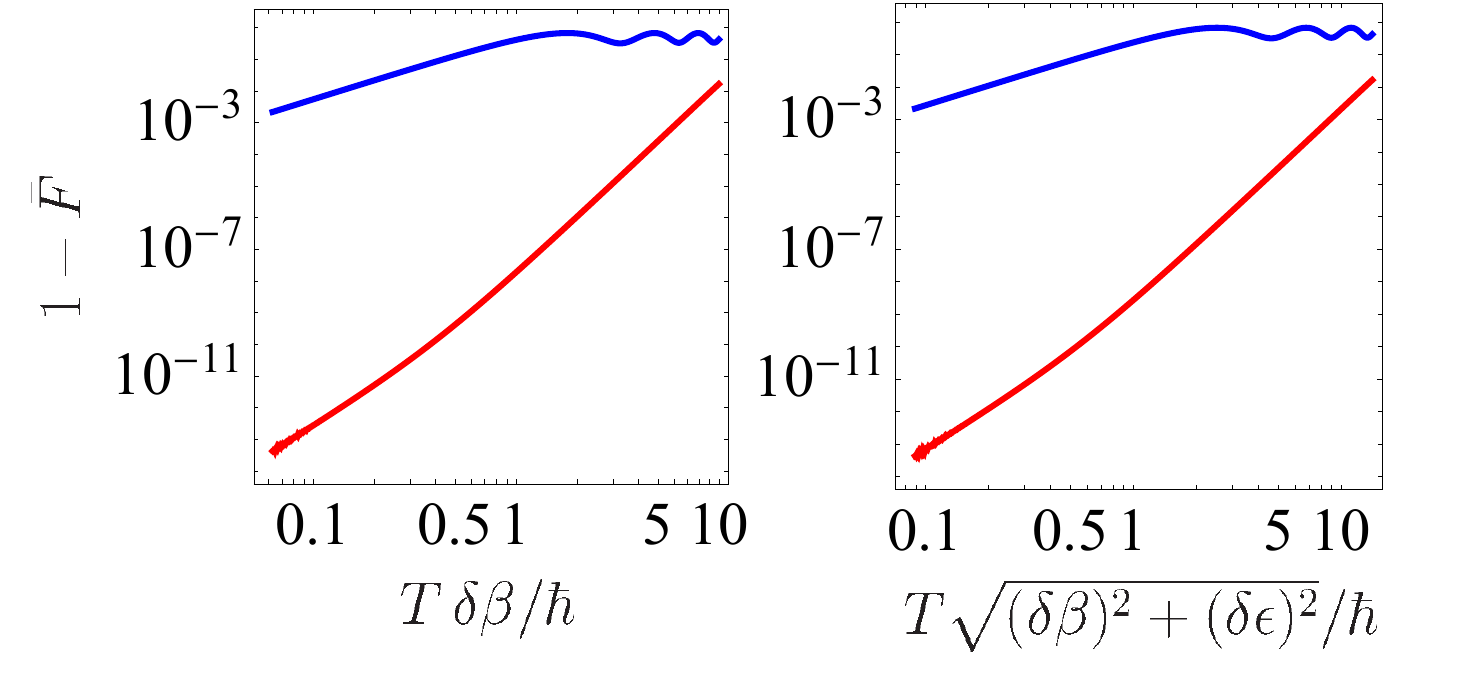}
		\caption{Plot of the infidelity in rotations by $\frac{\pi}{2}$ for $\delta\epsilon=0$ and as a function of $\delta\beta$
		(left) and for $\delta\epsilon=\delta\beta$ and as a function of $\sqrt{(\delta\beta)^2+(\delta\epsilon)^2}$ (right) for
		$f(\Omega)=\left (\frac{\Omega}{\Omega_0}\right )^{-3}$ and $\Omega_0=0.05\frac{\hbar}{T}$.  The blue curves represent the
		na\"ive pulse sequences, while the red curves represent the corrected sequences.}
	\label{Fig:IF_-3}
\end{figure}

\begin{figure}[htb]
	\centering
		\includegraphics[width=\columnwidth]{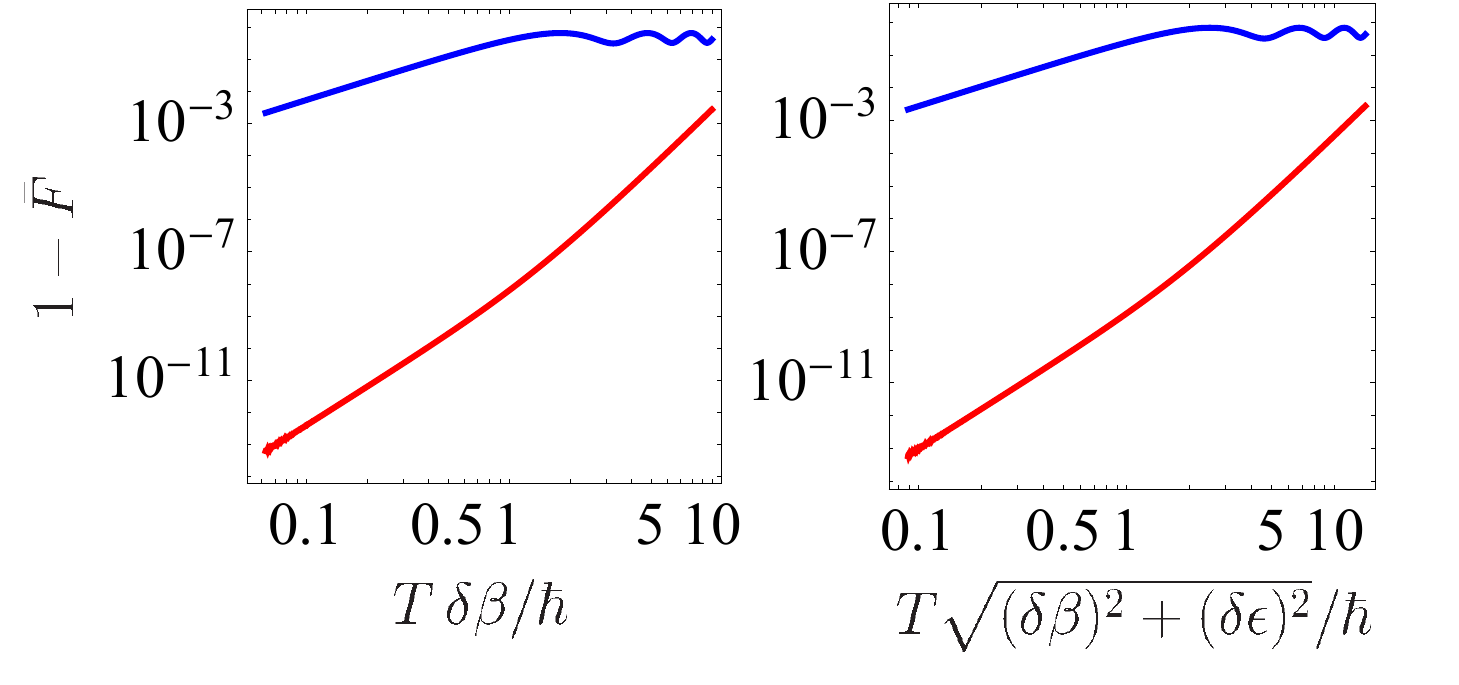}
		\caption{Plot of the infidelity in rotations by $\frac{\pi}{2}$ for $\delta\epsilon=0$ and as a function of $\delta\beta$
		(left) and for $\delta\epsilon=\delta\beta$ and as a function of $\sqrt{(\delta\beta)^2+(\delta\epsilon)^2}$ (right) for
		$f(\Omega)=\left (\frac{\Omega}{\Omega_0}\right )^{-1}$ and $\Omega_0=0.05\frac{\hbar}{T}$.  The blue curves represent the
		na\"ive pulse sequences, while the red curves represent the corrected sequences.}
	\label{Fig:IF_-1}
\end{figure}

\begin{figure}[htb]
	\centering
		\includegraphics[width=\columnwidth]{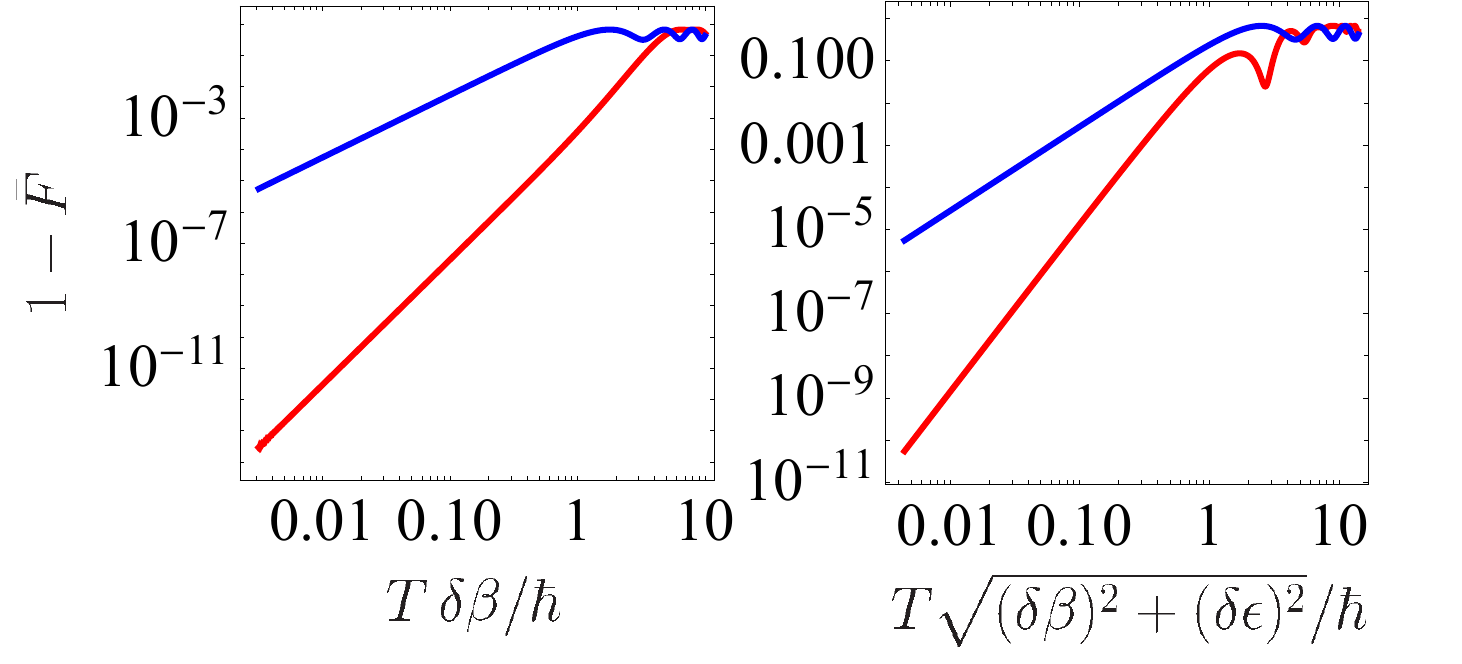}
		\caption{Plot of the infidelity in rotations by $\frac{\pi}{2}$ for $\delta\epsilon=0$ and as a function of $\delta\beta$
		(left) and for $\delta\epsilon=\delta\beta$ and as a function of $\sqrt{(\delta\beta)^2+(\delta\epsilon)^2}$ (right) for
		$f(\Omega)=\left (\frac{\Omega}{\Omega_0}\right )^3$ and $\Omega_0=8\frac{\hbar}{T}$.  The blue curves represent the na\"ive
		pulse sequences, while the red curves represent the corrected sequences.}
	\label{Fig:IF_3}
\end{figure}

\begin{figure}[htb]
	\centering
		\includegraphics[width=\columnwidth]{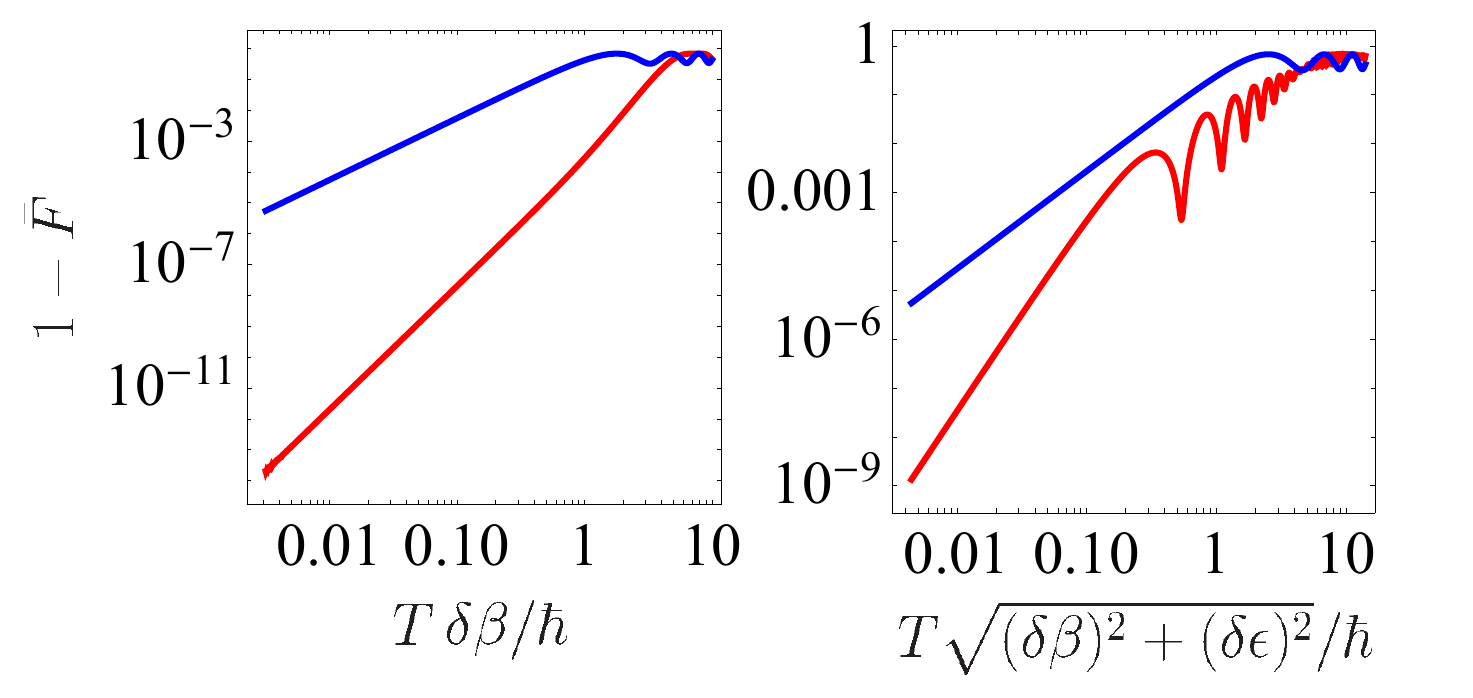}
		\caption{Plot of the infidelity in rotations by $\frac{\pi}{2}$ for $\delta\epsilon=0$ and as a function of $\delta\beta$
		(left) and for $\delta\epsilon=\delta\beta$ and as a function of $\sqrt{(\delta\beta)^2+(\delta\epsilon)^2}$ (right) for
		$f(\Omega)=\left (\frac{\Omega}{\Omega_0}\right )^5$ and $\Omega_0=8\frac{\hbar}{T}$.  The blue curves represent the na\"ive
		pulse sequences, while the red curves represent the corrected sequences.}
	\label{Fig:IF_5}
\end{figure}

We now evaluate the error in our error-corrected pulse sequences against the corresponding na\"ive operations,
i.e., the operation that, in the absence of noise, will perform the desired rotation; in this case, said operation
consists of just a single square pulse of appropriate magnitude and duration.  We will use the state-averaged
infidelity as our measure of the error.  The fidelity of a gate as a function of the initial state $\ket{\psi}$,
$F(\psi)$, is simply the probability that one will measure the qubit in the state that the gate was intended to
evolve it into\cite{NielsenBook}:
\begin{equation}
F(\psi)=\left |\bra{\psi}U^\dag R\ket{\psi}\right|^2,
\end{equation}
where $R$ is the ideal gate and $U$ is the actual gate.  We define the infidelity as $1-F(\psi)$.
The state-averaged infidelity $1-\bar{F}$ is simply the infidelity averaged over all distinct qubit states,
parametrized as
\begin{equation}
\ket{\psi(\theta,\varphi)}=\cos\left (\frac{\theta}{2}\right )\ket{0}+e^{i\varphi}\sin\left (\frac{\theta}{2}\right )\ket{1}.
\end{equation}
The state-averaged infidelity may be written as
\begin{equation}
1-\bar{F}=1-\frac{1}{4\pi}\int d\Omega\,F[\psi(\theta,\varphi)].
\end{equation}
Evaluating the integral, we obtain
\begin{equation}
1-\bar{F}=\tfrac{2}{3}-\tfrac{1}{6}[\mbox{Tr}(U^\dag R)]^2. \label{Eq:StateAvIF}
\end{equation}
We evaluate the infidelity numerically for rotations by $\frac{\pi}{2}$ and for $f(\Omega)\propto\Omega^{2n+1}$
with $n=-2$, $-1$, $1$, and $2$, and provide plots of the results for both the na\"ive and corrected sequences
for two cases, $\delta\epsilon=0$ and $\delta\epsilon=\delta\beta$ in Figs.\ \ref{Fig:IF_-3}--\ref{Fig:IF_5}.
Note that we do not present plots for $\delta\beta=0$, since, as argued earlier, the corrected sequences result
in {\it exact} cancellation of error, and thus the infidelity for these sequences would be exactly zero for all
values of $\delta\epsilon$.  We choose to present just the plots for rotations by $\frac{\pi}{2}$ because they
are representative of the general features found for all rotations.  We see that, indeed, the corrected pulse
sequences result in lower infidelities by several orders of magnitude than the na\"ive sequences, and we can see
that the scaling of the infidelity changes from second order to fourth order in $\delta\beta$ and $\delta\epsilon$, thus
showing that the leading-order effects of noise-induced errors are canceled exactly.  In some cases, the decrease
in infidelity is especially large; this is likely connected to the fact that, if $\delta\beta$ were to be zero,
then these sequences would exactly cancel noise-induced error.  We note that the infidelity, in some cases, oscillates
for larger values of $\delta\beta$.  A similar effect can be seen in the results of Ref.\ \onlinecite{WangNPJQI2015}
(which considered two-qubit gates), and is an interference effect.

\section{Trapezoidal pulses}
We now consider the case of trapezoidal pulses.  Such pulses are similar to the square pulses just considered, but include ramp
up or down segments as well.  We consider such a shape because, in reality, experimental setups cannot produce perfect square
pulses---there will always be a finite rise or fall time.  The basic building blocks of these sequences will be of the form,
\begin{equation}
\Omega(t)=
\begin{cases}
\frac{\Omega_n-\Omega_{n-1}}{\alpha T'}t+\Omega_{n-1}, & 0\leq t<\alpha T', \\
\Omega_n, & \alpha T'\leq t\leq T'.
\end{cases}
\end{equation}
This pulse represents a linear ramp up/down of the driving field from $\Omega_{n-1}$ to $\Omega_n$ over a time $\alpha T'$, where
$T'$ is the duration of this segment of the pulse train and $0<\alpha<1$ is an arbitrary constant determining what fraction of the
segment is spent on the ramp up/down, followed by holding the field at a constant value $\Omega_n$ for the remainder of the
duration.

We will build up our corrected pulse sequences from four of these segments, plus a final ramp down to zero.  We also start
the first segment from zero (i.e., ramp up from zero to a constant).  For simplicity, we will assume that the four segments
are all of equal length, and that all ramps up and down take the same amount of time (i.e., the four segments all have
duration $T'$, and all ramps up and down, including the final one, have duration $\alpha T'$).  We can express the formulas
giving the error correction conditions in analytic form as follows.  We first calculate the contributions to the left-hand
sides of these conditions for a single segment of the form given above.  Doing this, we find that the contributions to the
(ideal) rotation angle about the $z$ axis, to the left-hand side of Eq.\ \eqref{Eq:DECondition} and to the real and imaginary
parts of the left-hand side of Eq.\ \eqref{Eq:DBCondition} are, respectively,
\begin{widetext}
\begin{eqnarray}
\phi_n&=&\omega'_n-\tfrac{1}{2}(\omega'_n-\omega'_{n-1})\alpha, \\
\int_0^1[\omega'_n(\tau)]^k\,d\tau&=&\frac{(\omega'_{n})^{k+1}-(\omega'_{n-1})^{k+1}}{\omega'_n-\omega'_{n-1}}\frac{\alpha}{k+1}+(\omega')_n^k(1-\alpha), \\
\mbox{Re }\int_0^1 e^{-i\phi_n(\tau)}\,d\tau&=&\frac{2}{\omega'_n}\cos\left (\frac{\omega'_{n-1}\alpha+\omega'_n}{2}\right )\sin\left [\frac{\omega'_n(1-\alpha)}{2}\right ]+\sqrt{\frac{\pi\alpha}{|\omega'_n-\omega'_{n-1}|}}\left\{\cos\left (\frac{\alpha}{2}\frac{(\omega'_{n-1})^2}{|\omega'_n-\omega'_{n-1}|}\right )\right.\cr
&\times&\left [C\left (\sqrt{\frac{\alpha}{\pi}|\omega'_n-\omega'_{n-1}|}\frac{\omega'_n}{\omega'_n-\omega'_{n-1}}\right )-C\left (\sqrt{\frac{\alpha}{\pi}|\omega'_n-\omega'_{n-1}|}\frac{\omega'_{n-1}}{\omega'_n-\omega'_{n-1}}\right )\right ]+\cr
&+&\left.\sin\left (\frac{\alpha}{2}\frac{(\omega'_{n-1})^2}{|\omega'_n-\omega'_{n-1}|}\right )\left [S\left (\sqrt{\frac{\alpha}{\pi}|\omega'_n-\omega'_{n-1}|}\frac{\omega'_n}{\omega'_n-\omega'_{n-1}}\right )-S\left (\sqrt{\frac{\alpha}{\pi}|\omega'_n-\omega'_{n-1}|}\frac{\omega'_{n-1}}{\omega'_n-\omega'_{n-1}}\right )\right ]\right\}, \nonumber \\ \\
\mbox{Im }\int_0^1 e^{-i\phi_n(\tau)}\,d\tau&=&-\frac{2}{\omega'_n}\sin\left (\frac{\omega'_{n-1}\alpha+\omega'_n}{2}\right )\sin\left [\frac{\omega'_n(1-\alpha)}{2}\right ]+\mbox{sgn}(\omega'_n-\omega'_{n-1})\sqrt{\frac{\pi\alpha}{|\omega'_n-\omega'_{n-1}|}}\left\{\sin\left (\frac{\alpha}{2}\frac{(\omega'_{n-1})^2}{|\omega'_n-\omega'_{n-1}|}\right )\right.\cr
&\times&\left [C\left (\sqrt{\frac{\alpha}{\pi}|\omega'_n-\omega'_{n-1}|}\frac{\omega'_n}{\omega'_n-\omega'_{n-1}}\right )-C\left (\sqrt{\frac{\alpha}{\pi}|\omega'_n-\omega'_{n-1}|}\frac{\omega'_{n-1}}{\omega'_n-\omega'_{n-1}}\right )\right ]-\cr
&-&\left.\cos\left (\frac{\alpha}{2}\frac{(\omega'_{n-1})^2}{|\omega'_n-\omega'_{n-1}|}\right )\left [S\left (\sqrt{\frac{\alpha}{\pi}|\omega'_n-\omega'_{n-1}|}\frac{\omega'_n}{\omega'_n-\omega'_{n-1}}\right )-S\left (\sqrt{\frac{\alpha}{\pi}|\omega'_n-\omega'_{n-1}|}\frac{\omega'_{n-1}}{\omega'_n-\omega'_{n-1}}\right )\right ]\right\}, \nonumber \\
\end{eqnarray}
\end{widetext}
where $\tau=\frac{t}{T'}$, $\omega'_n=\frac{T'}{\hbar}\Omega_n$, and $C(x)$ and $S(x)$ are the Fresnel integrals,
\begin{eqnarray}
C(x)&=&\int_0^x\cos\left (\frac{\pi}{2}t^2\right )\,dt, \\
S(x)&=&\int_0^x\sin\left (\frac{\pi}{2}t^2\right )\,dt.
\end{eqnarray}
We also derive similar expressions for the final ramp down:
\begin{eqnarray}
\phi_f&=&\tfrac{1}{2}\omega'_f\alpha, \\
\int_0^1[\omega'_f(t)]^k\,d\tau&=&\frac{(\omega'_f)^k}{k+1}\alpha, \\
\mbox{Re }\int_0^1 e^{-i\phi_f(\tau)}\,d\tau&=&\sqrt{\frac{\pi\alpha}{|\omega'_f|}}\left [\cos\left (\frac{\alpha|\omega'_f|}{2}\right )C\left (\sqrt{\frac{\alpha}{\pi}|\omega'_f|}\right )\right.\cr
&+&\left.\sin\left (\frac{\alpha|\omega'_f|}{2}\right )S\left (\sqrt{\frac{\alpha}{\pi}|\omega'_f|}\right )\right ], \\
\mbox{Im }\int_0^1 e^{-i\phi_f(\tau)}\,d\tau&=&\mbox{sgn}(\omega'_f)\sqrt{\frac{\pi\alpha}{|\omega'_f|}}\left [\sin\left (\frac{\alpha|\omega'_f|}{2}\right )\times\right.\cr
\times C\left (\sqrt{\frac{\alpha}{\pi}|\omega'_f|}\right )&-&\left.\cos\left (\frac{\alpha|\omega'_f|}{2}\right )S\left (\sqrt{\frac{\alpha}{\pi}|\omega'_f|}\right )\right ],
\end{eqnarray}
where $\omega'_f=\frac{T'}{\hbar}\Omega_f$ is the driving field at the beginning of this final ramp down.  In order
to obtain the error cancellation conditions, we can simply add the contributions given above to the intended rotation
angle and to the driving field noise cancellation condition.  However, one cannot simply add together the contributions
to the $\delta\beta$ cancellation conditions.  The contributions to the overall rotation angle and to the left-hand
side of Eq.\ \eqref{Eq:DECondition} only involve a simple integral of a function of $\Omega(t)$---$\Omega(t)$ itself 
and $[\Omega(t)]^k$, respectively, to be exact---and thus the contribution of each segment simply adds to that of the
previous segment(s).  However, the contributions to the left-hand side of Eq.\ \eqref{Eq:DBCondition} are more complicated---they
involve an integral of a function that itself involves an integral of $\Omega(t)$, namely, $\exp\left [-\frac{i}{\hbar}\int_0^t\Omega(t)\,dt\right ]$.
We may still use the above expressions to build up the full error cancellation conditions by noting that the contribution
from the $n^\text{th}$ segment acquires an extra phase factor of $\exp\left (-i\sum_{j=1}^{n-1}\phi_j\right )$ in the
full expression for the $\delta\beta$ error cancellation condition.

As before, we solve the resulting equations numerically.  We consider here the case, $f(\Omega)\propto\Omega^3$ and
$\alpha=0.05$.  We find that, once again, it is possible to correct rotations by arbitrary angles, and we give the
parameters that we obtain for certain rotations in Table \ref{Tab:PulseParameters_Trapezoidal_1}.  It is straightforward,
if tedious, to obtain similar parameters for other powers (i.e., values of $k$ different from $3$).
\begin{table}[htb]
	\centering
		\begin{tabular}{c|c|c|c|c}
			$\phi$ & $\omega'_1$ & $\omega'_2$ & $\omega'_3$ & $\omega'_4$ \\
			\hline
			\hline
			$\pi/4$ & 2.36859 & -4.34927 & -1.39592 & 4.16199 \\
			\hline
			$\pi/2$ & 2.66739 & -4.73666 & -0.80657 & 4.44664 \\
			\hline
			$3\pi/4$ & 2.9033 & -5.03395 & -0.197971 & 4.68482 \\
			\hline
			$\pi$ & 3.00887 & -5.26353 & 0.495858 & 4.90039 \\
			\hline
			$5\pi/4$ & 2.9789 & -5.47004 & 1.30658 & 5.11155 \\
			\hline
			$3\pi/2$ & 2.96033 & -5.73813 & 2.17134 & 5.31884 \\
			\hline
			$7\pi/4$ & 3.14526 & -6.10634 & 2.94228 & 5.51659 \\
			\hline
			$2\pi$ & 3.52834 & -6.53513 & 3.59846 & 5.69151
		\end{tabular}
		\caption{Pulse parameters for different rotation angles $\phi$ for $f(\Omega)\propto\Omega^3$ and using the
		trapezoidal pulses with $\alpha=0.05$.}
	\label{Tab:PulseParameters_Trapezoidal_1}
\end{table}

\subsection{Evaluation of error}
\begin{figure}[htb]
	\centering
		\includegraphics[width=\columnwidth]{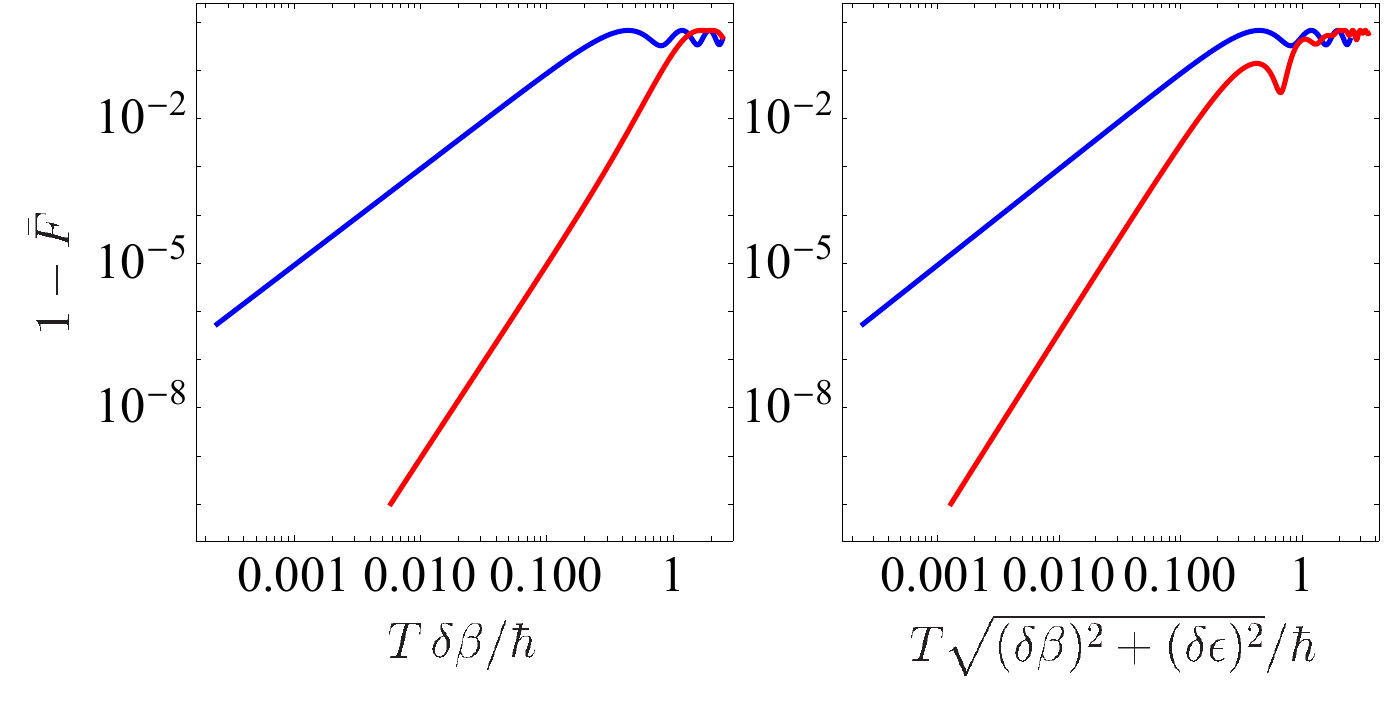}
		\caption{Plot of the infidelity in rotations by $\frac{\pi}{2}$ for $\delta\epsilon=0$ and as a function of $\delta\beta$
		(left) and for $\delta\epsilon=\delta\beta$ and as a function of $\sqrt{(\delta\beta)^2+(\delta\epsilon)^2}$ (right) for
		the trapezoidal pulse sequences with $\alpha=0.05$.  Here, we take $f(\Omega)=\left (\frac{\Omega}{\Omega_0}\right )^3$
		and $\Omega_0=1.975\frac{\hbar}{T'}$.  The blue curves represent the na\"ive pulse sequences, while the red curves
		represent the corrected sequences.}
	\label{Fig:IF_3_TrapPulses}
\end{figure}

We now evaluate these pulse sequences to show that they indeed cancel noise-induced errors to first order.  We once
again adopt the state-averaged infidelity, Eq.\ \eqref{Eq:StateAvIF}, as our metric.  In this case, we cannot solve
for the evolution operator for the ramp up/down portions of the pulse sequences analytically, so we must do so
numerically.  We use as our ``na\"ive'' pulse sequence a single trapezoidal pulse that begins and ends with $\Omega=0$
and has the same duration as our corrected pulse sequence.  We determine the infidelity for the same two cases as
for the rectangular pulses, $\delta\epsilon=0$ and $\delta\epsilon=\delta\beta$, and plot our results in Fig.\ \ref{Fig:IF_3_TrapPulses}.
We see that, once again, the pulse sequences work as intended---they cancel the effect of noise-induced errors at leading
order, and in fact reduce infidelity by many orders of magnitude if the noise is sufficiently small.

\section{Conclusion}
We have determined conditions under which it is possible to cancel both (longitudinal) driving field noise and (transverse)
orthogonal-to-driving-field noise in a qubit driven along its logical $z$ axis.  We considered the Hamiltonian, Eq.\ \eqref{Eq:Hamiltonian},
which we repeat here for convenience:
\begin{equation}
H=\tfrac{1}{2}[\Omega(t)+f(\Omega)\,\delta\epsilon]\sigma_z+\delta\beta\,\sigma_x,
\end{equation}
where $f(\Omega)$ is some function of the driving field $\Omega(t)$.  We specifically set out to determine what forms of this
function allow us to cancel the effects of the noise terms, the driving field noise ($\delta\epsilon$) and orthogonal-to-driving-field
noise ($\delta\beta$).  We considered power-law forms for $f(\Omega)$, i.e., $f(\Omega)\propto\Omega^k$.
We began by deriving the conditions for canceling the effects of these two error terms to first order.  We showed that the
condition for canceling the effects of the $\delta\beta$ term to first order was identical to that obtained in Ref.\ \onlinecite{ZengNJP2018}
in the case that this was the only term present.  Therefore, the corresponding condition for the $\delta\epsilon$ term simply
imposes a further constraint on the driving field.  This second constraint, unfortunately, does not permit cancellation of
driving field noise-induced errors in arbitrary rotations for the power-law form of $f(\Omega)\propto\Omega^k$ if the exponent
$k=0$ or $1$.  More specifically, cancellation of these errors is only possible for the identity operation if $k=1$, and is
not possible for {\it any} operations if $k=0$.  Furthermore, it does not allow for cancellation if $k$ is any even integer.
However, we find that, if $k$ is any odd integer besides $1$, then we {\it can} correct noise-induced errors in the presence
of both driving field and orthogonal-to-driving-field noise.  We in fact determine four-part sequences of square pulses of equal
duration (corresponding to the number of real-valued constraints that we must satisfy) that do just that.  Our choice of square
pulses is motivated by the work of Ref.\ \onlinecite{ZengArXiv}, which shows that, for the system that we consider, square pulses
minimize the time duration of all possible error-correcting pulses.  We then evaluate the effectiveness of these sequences at
combating errors by determining the infidelity of both the na\"ive and corrected versions of sequences for implementing rotations
by $\frac{\pi}{2}$.

We should emphasize that the condition under which error correction is possible only applies to the system that we considered,
in which the only intentionally applied field is the driving field along the logical $z$ axis, i.e., there are no fields applied
along, say, the logical $x$ axis.  For example, all of the work on \textsc{supcode} and its generalizations\cite{WangNatComm2012,WangPRA2014,ThrockmortonPRB2017},
which considers a system in which there are driving fields along two axes (specifically a singlet-triplet qubit), i.e.,
\begin{equation}
H=\tfrac{1}{2}[\Omega(t)+f(\Omega)\,\delta\epsilon]\sigma_z+(\beta+\delta\beta)\sigma_x,
\end{equation}
assumes that $f(\Omega)\propto\Omega$, which we have just shown does not permit error cancellation for the Hamiltonian considered
in this work.  In fact, the specific platform being considered in those works is restricted to positive values of $\Omega(t)$, and
thus error correction should be possible for any power-law form of $f(\Omega)$.  An important extension of the present work and of
the work that inspired it\cite{ZengNJP2018,ZengArXiv} would be to consider the case in which we include an intentionally applied
field along the logical $x$ axis.  This, however, would be a much more difficult problem to solve, as there is no known exact solution
for arbitrary pulse shapes to the resulting Hamiltonian.  This remains an important future problem to solve, but our work shows that,
at least when the driving field is only along the logical $\sigma_z$ direction, geometrical dynamical decoupling techniques can
indeed correct for errors arising from quasistatic noise existing along both $x$ and $z$ directions provided certain well-defined
conditions are satisfied by the longitudinal noise term.  Our results apply to all physical qubits obeying Eq.\ \eqref{Eq:Hamiltonian}.

\acknowledgments
This work is supported by the Laboratory for Physical Sciences.

\end{document}